\documentclass[11pt,twoside]{article}
\usepackage{asp2006}
\usepackage{txfonts}
\usepackage{graphicx}
\begin{document}
\title{Low frequency results for the radio galaxies and the role of GMRT}
\author{Dharam~Vir~Lal}
\affil{MPI f\"ur Radioastronomie,
      Auf dem H\"ugel 69,
      53121 Bonn, Germany.}
\begin{abstract}
We have studied several radio galaxies at low radio frequencies using GMRT.
Our prime motivation to detect faint radio
emission at very low frequencies due to low energy electrons.

Our results provide evidence that there exists two classes of sources on
morphological grounds.
The first class is explained by the simple picture of spectral electron ageing
but in the second class the low-frequency
synchrotron emission fades \mbox{(nearly) as} rapidly as high-frequency synchrotron
emission.
In addition, in several sources, the spectra of low-surface-brightness
features are flatter than the spectra of high-surface-brightness features,
which suggests that either the simple picture of spectral
electron ageing needs revision or we need to
\mbox {re-examine the formation mechanism of such sources.}

The images and statistics, and the relevance of these results
along with the role of GMRT in exploring several unknowns are presented.
\end{abstract}

\section{Introduction}

Radio galaxies display a wide
range of structures in radio images. The most common large-scale structures are
called lobes: these are double, often fairly symmetrical, roughly ellipsoidal
structures placed on either side of the active nucleus. A significant minority
of low-luminosity sources exhibit structures usually known as plumes which are
much more elongated. Some radio galaxies show one or two long narrow features
known as jets coming directly from the nucleus and going to the lobes.
Hence, the radio galaxies morphologically are divided into FR\,I and FR\,II types.
The FR\,I objects (`edge-darkened doubles') generally contain
prominent, often two-sided jets, with the lobe emission trailing off into
intergalactic space.  Whereas, the FR\,II objects (`edge-brightened doubles',
`classical doubles') contain hot spots in one or both lobes.

It seems that FR\,II radio galaxies (see Fig.~\ref{fig:3c98}) are usually
isolated field sources and form a remarkably homogeneous set.
On the other hand, the FR\,I radio galaxies are much more varied.  They
are not only found as isolated field sources, they are also often found
in clusters \citep{Perley1989}. For example, head-tail galaxies,
which are FR\,I objects, are almost always associated with clusters of galaxies and
are characterized by a highly elongated radio structure with the associated optical
galaxy at one end.
A classical example (see Fig.~\ref{fig:3c129}) is 3C\,129 (a narrow-angle-tail)
which, along with its companion 3C\,129.1 (a wide-angle-tail), is a member of
an X-ray cluster.
In the canonical picture, the characteristic shape is explained by the
kinematics of the source, which is governed by the dominant gravitational force
in the cluster and the properties of the beams and jets \citep{2004AandA}.
Similarly, wide-angle-tail galaxies are intermediate between standard FR\,I and
FR\,II objects, with collimated jets and sometimes hotspots, but with plumes
rather than lobes, found at or near the centres of clusters.  Here again,
the characteristic shape is mainly due to the dominant gravitational force in
the cluster.

Therefore, the general morphology of radio sources lends encouragement to ideas
of confinement, or at least radio morphology partly governed by an external gas
(\citealt{Hard2005}, and see Hardcastle in this proceeding), but some objects
persist in complicating the simple picture.
For example, B1059$+$169 (Fig. \ref{fig:b1059}) which shows X-shaped structure.
Such sources are characterized by two axes, the `wing' axis oriented
at an angle to the `active' axis, giving the total source an `X' shape.
These two sets of lobes usually pass symmetrically through the centre of the
associated host galaxy, and the majority of these sources are FR\,II and the
rest are either FR\,I or mixed.
These have been modelled via backflow from the lobes along with
buoyancy-driven outflow.
Precession of the radio axis also comes to mind.
They have also been put forth as derivatives of central engines that have been
reoriented, perhaps due to a minor merger.
Alternatively, they may also result from two pairs of jets that are
associated with a pair of unresolved AGNs (see\ \citealt{2007mnras} for a
detailed summary).
The key point is that interpretation of radio source morphology, in general,
is not necessarily simple.

\medskip
\noindent
{\bf {Spectral ageing}}
Under the standard model of a FR\,II source evolution, the source grows longer
as the jet pushes back the external medium.
If there s no significant re-acceleration within the lobes and no significant
transport of particles since the last acceleration, then the central regions
are `older' than the outer regions, in the sense that particles located in the
central regions were processed through the hot spot shock at a time before
those currently radiating at the ends of the radio source.
Hence, the spectral index of the radiation from the central regions should have
a steeper spectral index than that at the ends.
Whereas, in the case of FR\,I sources, e.g., in almost all head-tail and
wide-angle-tail galaxies, the radio spectrum has been found to steepen with
distance along the tail and has been interpreted in terms of ageing of electron
population.

Finally, in several of the formation scenarios mentioned above for X-shaped
sources, the wings are interpreted as relics of past radio jets and the active
lobes as the newer ones.  Hence, the wings are expected to show steeper spectra
than the active lobes in standard models for electron energy evolution.

\section{Observations and Data Analysis}

In recent years we have used the GMRT (\citealt{Rao2002}) to study morphology
of several radio galaxies at 240~MHz and 610~MHz, with a resolution of about
10~arcsec~and~5~arcsec, respectively.
The GMRT visibility data were analyzed using {\sc aips} in the standard manner.
All details of the analysis procedures used are given in \citet{2008mnras}.

The GMRT has a hybrid configuration with 14 of its 30 antennae located in a
central compact array with a size ∼1.1 km (comparable to the VLA D
configuration) and the remaining antennae distributed in a roughly `Y'-shaped
configuration, giving a maximum baseline length of ∼25 km (comparable to the
VLA B configuration).
Hence, a single observation with the GMRT samples the  ($u, v$)  plane on both
short and long baselines, and can map detailed source structure with a
reasonably good sensitivity.

\section{Results}

I bring together two remarkable results obtained using GMRT.

\begin{figure}
\begin{center}
\begin{tabular}{ll}
\includegraphics[width=6.0cm]{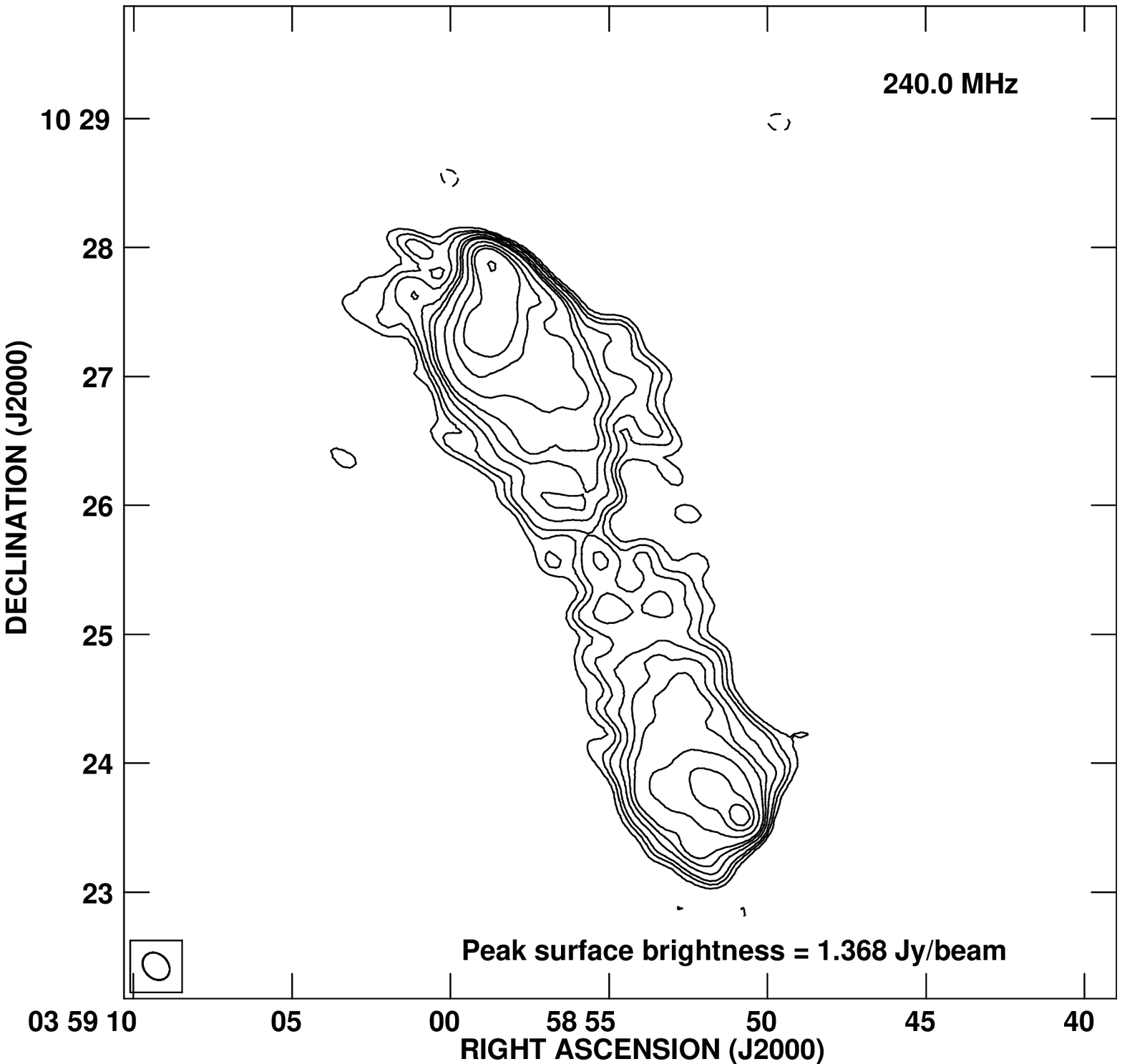} &
\includegraphics[width=6.0cm]{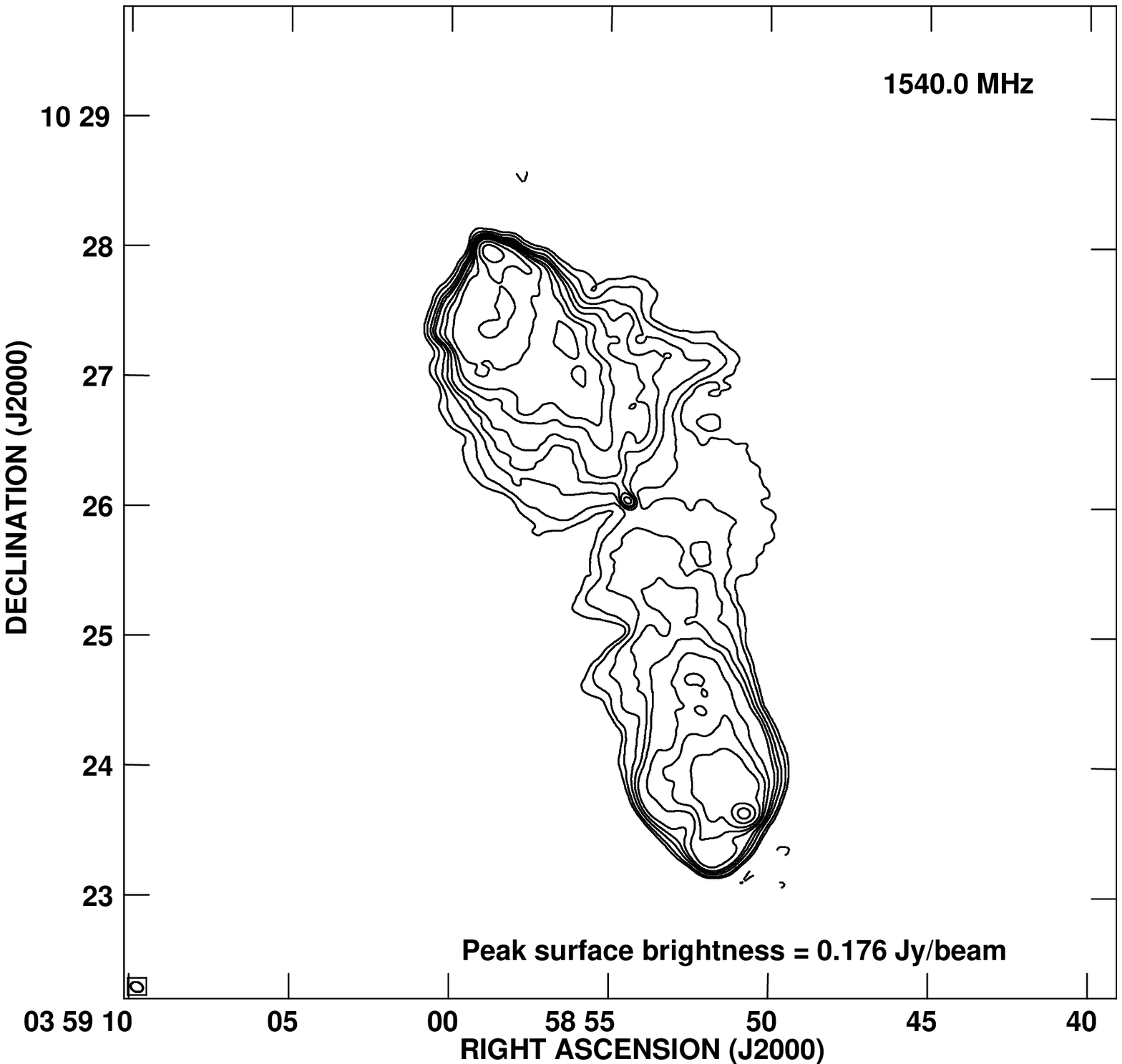} \\ [-0.58cm]
\end{tabular}
\end{center}
\caption{GMRT map of 3C\,98 at 240~MHz (left panel) and
the 1.5 GHz map (right panel) from the
Atlas catalog.  (Available at http://www.jb.man.ac.uk/atlas/~.)
\label{fig:3c98}}
\end{figure}

\begin{figure}
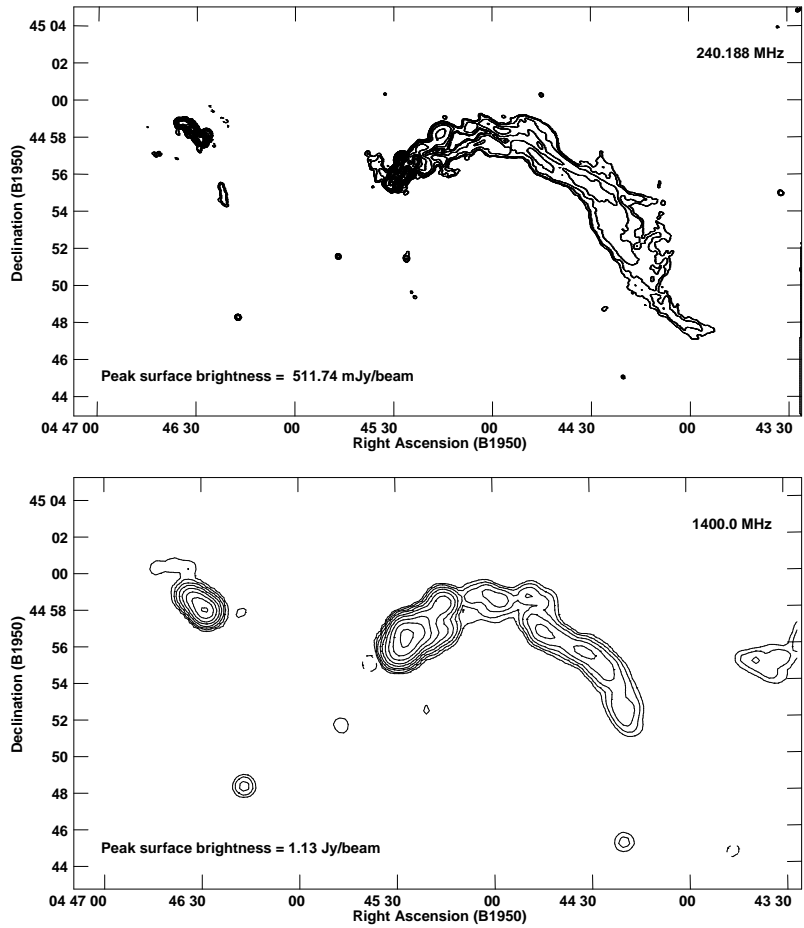

\begin{center}
\begin{tabular}{ll}
\includegraphics[angle=-90,width=10.9cm]{3c129_240.ps} \\ [-1.2cm]
\includegraphics[angle=-90,width=10.9cm]{3C129_NVSS.PS} \\ [-1.0cm]
\end{tabular}
\end{center}
\caption{Full synthesis GMRT map of 3C\,129 at 240~MHz (upper panel) showing
larger projected angular size than the NVSS map at 1.4 GHz (lower panel).
\label{fig:3c129}}
\end{figure}

\medskip
\noindent
{\bf {Similar/Dis-similar radio morphologies}}
It is remarkably unusual that any FR\,II source has a morphology at low
radio frequency (less than a few hundred MHz) that is different from its high
radio frequency (GHz) morphology.
Fig.~\ref{fig:3c98} shows 3C\,98 as an example showing similar radio
morphologies at 240 MHz and 1.5~GHz. In addition the 74~MHz and the 4.9~GHz
radio morphologies are also identical.
The observations of some 3C radio galaxies at 151\,MHz and 1.4\,GHz
\citep{Leahy1989} show that the lobe lengths at these different frequencies are
the same and they also suggest that the particles responsible for the
low-frequency emission are entirely co-spatial with those responsible for the
high-frequency emission.
This suggests that synchrotron-emitting particles of all energies permeate the
lobe magnetic field in the same way, despite the fact that the higher energy
particles have shorter radiative lifetimes than the lower energy ones.
This hints that low-frequency synchrotron emission fades (nearly) as rapidly as
high-frequency synchrotron emission \citep{Blundell2008}; and so, are the
observations really inconsistent with the idea that synchrotron cooling
is not the dominant energy-loss mechanism for the synchrotron plasma?
\citet{HL2008} note that by the time we get to 100 GHz there really are fairly
obvious morphological differences in lobes, although the sensitivity of those
observations is not great.
Whereas, the FR\,I sources, in particular the head-tail galaxies and
wide-angle-tail galaxies show signs of synchrotron cooling in spectral index
images \citep{2004AandA} made using narrow frequency spacing (240 MHz and 1.4
GHz) and often show the presence of steep spectrum diffuse emission at low
radio frequencies, which is not seen at high radio frequencies
(Fig.~\ref{fig:3c129}).

Nevertheless, in the light of above data, our results provide evidence that there
exists two class of sources linked to FR\,I/FR\,II sources on morphological
grounds.
In one, where the low frequency radio images show morphologies that are
dis-similar to the morphologies at high frequencies, consistent with the simple
picture of spectral electron ageing.
In the other, where the low frequency radio images show morphologies that are similar to
the morphologies at high frequencies, which, possibly, suggests that the
simple picture of spectral ageing needs revision.
In more detail, it seems that along with the simple FR\,I/FR\,II
division, which depends on host-galaxy environment in the sense that the
FR\,I/FR\,II transition appears at higher luminosities in more massive
galaxies, the role played by the intracluster gas in deciding the source
morphology is also important.  In particular, the role of dominant
gravitational force in the cluster in defining the source morphology cannot be
ignored, which possibly is
key to the physical origin of the above two classes of sources.

\begin{figure}
\begin{center}
\begin{tabular}{ll}
\includegraphics[width=6.0cm]{B1059_169_240_PUB.PS} &
\includegraphics[width=6.0cm]{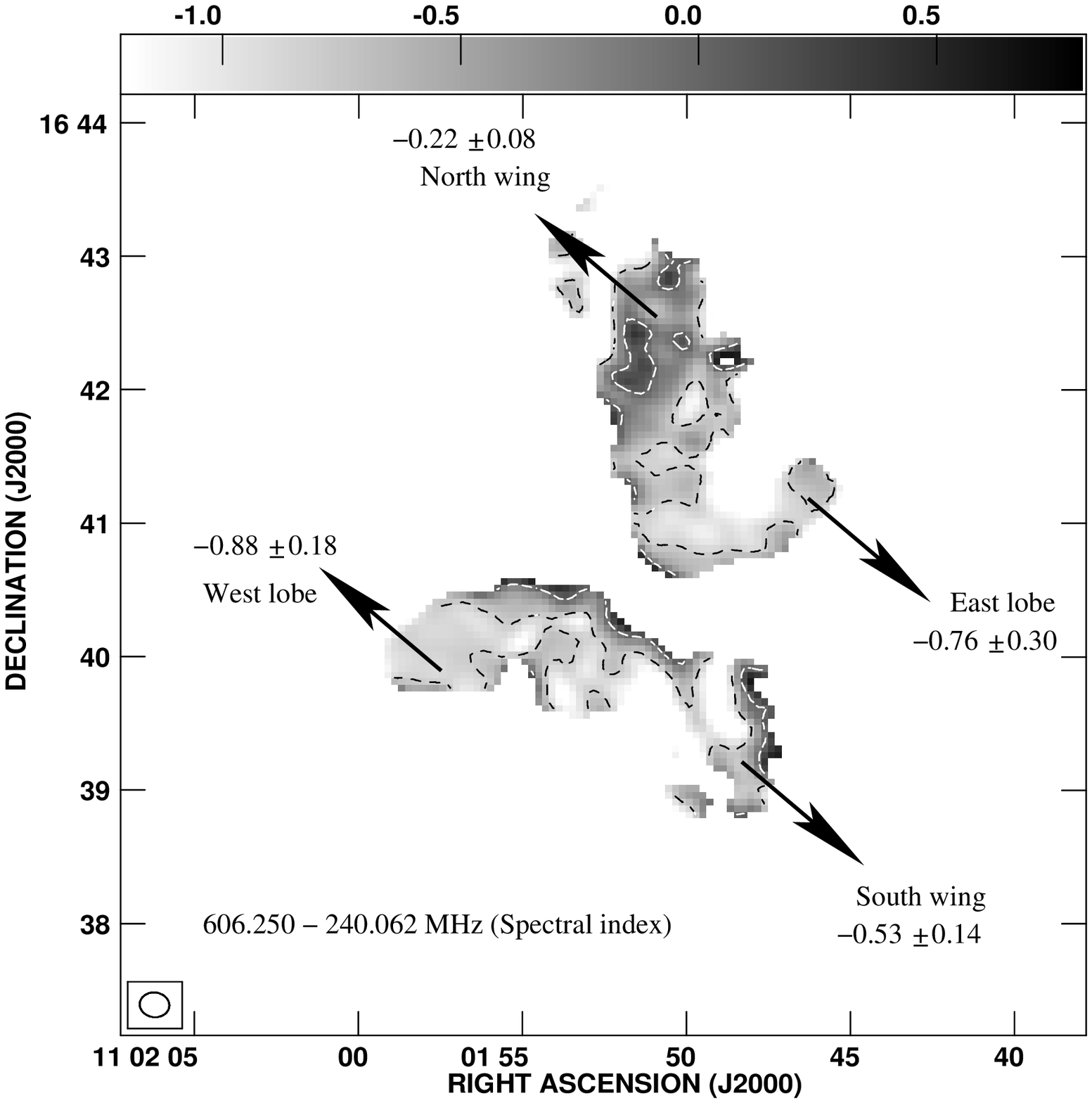} \\ [-1.0cm]
\end{tabular}
\end{center}
\caption{GMRT map of B1059$+$169 at 240~MHz (left panel) and
the distribution of the spectral index
between 240 MHz and 610 MHz (right panel),
where $S_\nu \propto \nu^\alpha$.
\label{fig:b1059}}
\end{figure}

\medskip
\noindent
{\bf {Low surface brightness features having flat spectra}}
Fig.~\ref{fig:b1059} shows B1059$+$169, a X-shaped
radio source seen in a cluster environment (Abell 1145), which is the dominant
radio galaxy and is ∼5.5 arcmin away from the cluster centre.  A companion is
detected on the 2MASS, coincident with the cluster centre.
Although the source is found in the cluster environment, surprisingly, its
spectral index map is unusual, that is, the wings have relatively flatter
spectral index compared to the active lobes.
Therefore, we remark that whenever there is a low-surface brightness feature
attached to a radio galaxy, it is not necessary that it should be a steep
spectrum feature.

In addition, if our understanding of synchrotron spectral ageing
at low radio frequencies is correct, then the spectral index results do not
favour several of the formation models for the X-shaped sources.
One plausible model is the \citet{2005mnras,2007mnras} model in which
X-shaped sources consist of two pairs of jets, which are associated with two
unresolved AGN.
However, exceptions exist, e.g., in the case of 3C\,403, Kraft et al. (2005)
favoured the hydrodynamic model.

\section{Looking Ahead}

Presently the GMRT observations have provided several high resolution, high
sensitivity images, and it is now possible to probe, statistically, the
morphological properties of radio sources.
I have presented above two unusual results, but it will still be important
to obtain observations using GMRT of complete, unbiased
samples of radio galaxies to draw general conclusions about the population
of radio galaxies as a whole.

\acknowledgements
We thank the staff of the GMRT that made these observations possible. GMRT is
run by the NCRA of the TIFR.
I warmly thank my collaborators M.J. Hardcastle, R.P. Kraft and A.P. Rao 
and especially A.L. Roy for a careful reading of this manuscript.


\end{document}